\documentclass[conference,10pt]{IEEEtran}
\usepackage{graphicx}
\usepackage{epstopdf}
\usepackage{amssymb}
\usepackage{amsmath,amsthm}
\usepackage{comment}
\usepackage{longtable}
\usepackage{multirow}
\usepackage{booktabs}
\usepackage{pslatex}
\usepackage{tabularx}
\usepackage{microtype}
\usepackage{blindtext, graphicx}
\usepackage{tikz}
\usetikzlibrary{positioning}

\renewcommand{\figurename}{Figure}

\def\ie{{i.e.},~}

\usepackage[all]{nowidow}

\usepackage{tabularx}

\newcommand{\revision}[1]{\textcolor{red}{\textbf{#1}}}
\renewcommand{\revision}[1]{{#1}}

\makeatletter
\newcommand{\rmnum}[1]{\romannumeral #1}
\newcommand{\Rmnum}[1]{\expandafter\@slowromancap\romannumeral #1@}
\makeatother

\usepackage{graphics}
\graphicspath{ {images/} }

\usepackage{amssymb,amsfonts}
\usepackage{amsmath}
\usepackage{amsthm}
\newtheorem{definition}{Definition}
\newtheorem{theorem}{Theorem}

\usepackage{algorithm}
\usepackage{algpseudocode}
\newcommand\Algphase[1]{%
\vspace*{-.7\baselineskip}\Statex\hspace*{\dimexpr-\algorithmicindent-2pt\relax}\rule{\textwidth}{0.2pt}%
\Statex\hspace*{-\algorithmicindent}\textbf{#1}%
\vspace*{-.7\baselineskip}\Statex\hspace*{\dimexpr-\algorithmicindent-2pt\relax}\rule{\textwidth}{0.2pt}%
}

\makeatletter
\def\BState{\State\hskip-\ALG@thistlm}
\makeatother

\IEEEoverridecommandlockouts
\IEEEpubid{\makebox[\columnwidth]{ 978-1-5386-1027-5/17/\$31.00~\copyright2017
IEEE \hfill} \hspace{\columnsep}\makebox[\columnwidth]{ }}

\usepackage{url}
\usepackage{cite}
\begin{document}

\title{Achieving Secure and Differentially Private Computations in Multiparty Settings}
\author{ 
    \IEEEauthorblockN{
    Abbas Acar\IEEEauthorrefmark{1},  
    Z. Berkay Celik\IEEEauthorrefmark{2},  
    Hidayet Aksu\IEEEauthorrefmark{1},  
    A. Selcuk Uluagac\IEEEauthorrefmark{1}, 
    Patrick McDaniel\IEEEauthorrefmark{2}}\\
    \IEEEauthorblockA{
    \IEEEauthorrefmark{1}CPS Security Lab, Department of ECE, Florida International University
    \\\{aacar001,haksu,suluagac\}@fiu.edu}
    \IEEEauthorblockA{\IEEEauthorrefmark{2}SIIS Laboratory, Department of CSE, The Pennsylvania State University
    \\\{zbc102, mcdaniel\}@cse.psu.edu}
}
\maketitle

\pagestyle{plain}

\begin{abstract}
Sharing and working on sensitive data in distributed settings from healthcare to finance is a major challenge due to security and privacy concerns. Secure multiparty computation (SMC) is a viable panacea for this, allowing distributed parties to make computations while the parties learn nothing about their data, but the final result. Although SMC is instrumental in such distributed settings, it does not provide any guarantees not to leak any information about individuals to adversaries. Differential privacy (DP) can be utilized to address this; however, achieving SMC with DP is not a trivial task, either. In this 
paper,
we propose a novel Secure Multiparty Distributed Differentially Private (SM-DDP) protocol to achieve secure and 
private computations in a multiparty environment. Specifically, with our protocol, we simultaneously
achieve SMC and DP in distributed settings focusing on linear regression on horizontally distributed data. That is, parties do not see each others'
data
and further, can not infer
information about
individuals
from the final constructed statistical model. Any statistical model function that allows independent calculation of local statistics can be computed through our protocol. The protocol 
implements
\textit{homomorphic encryption} for SMC and \textit{functional mechanism} for DP to achieve the desired security and privacy guarantees. In this work, we first introduce the theoretical foundation for the SM-DDP protocol and then evaluate its efficacy and performance on two different datasets. Our results show that one can achieve individual-level privacy 
through the
proposed protocol with distributed DP, which is independently applied by each party in a distributed fashion.
Moreover, our results
also
show that the SM-DDP protocol incurs minimal computational overhead, is scalable, and provides security and privacy guarantees.
\end{abstract}


\IEEEpeerreviewmaketitle

\section{Introduction}


Secure and private computation of statistical 
models is increasingly used in different operational settings from healthcare~\cite{kimura2016evaluation,celikPatient,DBLP:journals/corr/CelikAASUM17} to finance~\cite{bogdanov2012deploying} and security sensitive applications~\cite{freudiger2015controlled,celikForensic}. 
Given the distributed nature of these applications, security and privacy are mostly achieved by utilizing Secure Multiparty Computation (SMC). 
SMC allows distributed parties to jointly compute an agreed function over their private inputs without revealing those inputs to other parties. Each party learns the final result, but no other information.  
However, SMC has a major privacy concern for a targeted individual 
as it does not guarantee that the final result of distributed computation would not leak any information about an 
individual in a sensitive dataset. 
Privacy of individuals and their data can be easily violated.~\cite{el2013secure,narayanan2008robust,ganta2008composition}. 
Therefore, there is a need for a mechanism, where individual parties do not see each others' inputs and further can not infer their data from the final constructed model.
Indeed, combining SMC with Differential Privacy (DP) could solve this privacy problem as DP introduces sufficient noise into the final result to prevent any leakage 
about a single individual. 

\begin{figure}[t!]
    \centering
          \includegraphics[width=.5\textwidth,trim= 0cm 2cm 0cm 0cm]{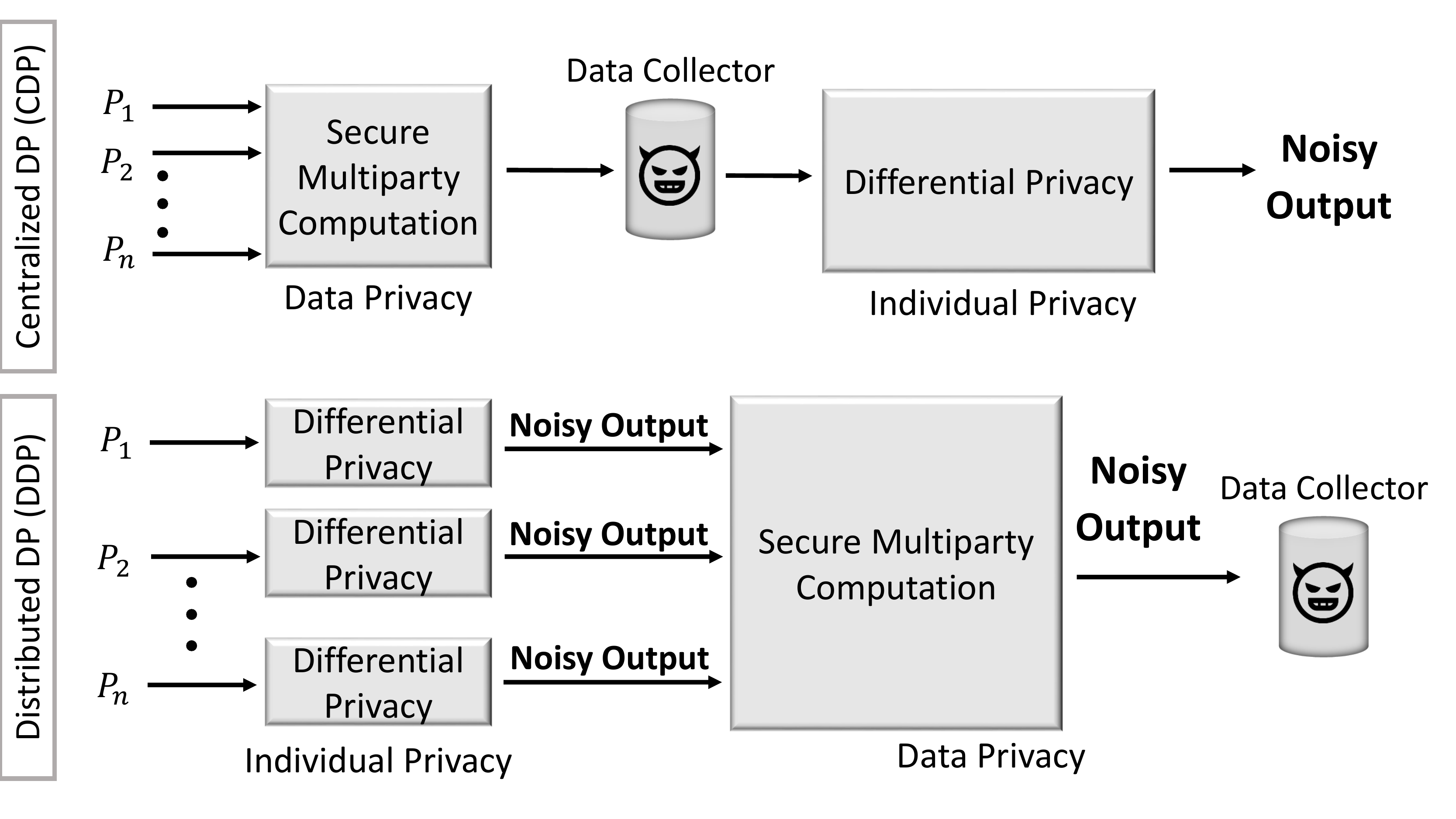}
          \caption{Illustration of secure multiparty computation with distributed and centralized differential privacy methods.}
\label{fig:com}
\end{figure}

However, combining SMC with DP is not a trivial task. In \revision{an} ideal case, a trusted data collector\footnote{A data collector is either one of the parties or a third party. Every discussion here applies to both of the types.} can collect the data, aggregate them and
add calibrated noise to the results of the queries (predictions) (Centralized DP (CDP) in Fig.~\ref{fig:com}). However, a trusted party does not exist in many real life scenarios. This technique would easily leak the model of the sensitive data to an untrusted data collector who collects the final model of the data. 
Even for scenarios with a trusted data collector, relying on the centralized entity makes it a single point of failure for the entire data collection mechanism. 

On the other hand, another mechanism involves applying 
a data sanitization technique (Distributed DP (DDP) in Fig.~\ref{fig:com}) directly on the local data held by the parties. In this case, the untrusted data collector can not infer individuals' data since sufficient noise is injected by DP to hide the individuals' data. However, this mechanism requires a meticulous analysis since it may lead to a divergent or excessive amount of accumulated noise due to DP at the data collector end. 
As such, this process may lead to a significant accuracy loss in the final models, which may cause catastrophic consequences in, for example, the healthcare domain. Therefore, enabling distributed differential privacy on local data with differential privacy guarantees on final results is a challenging problem.

In this paper, we are motivated to provide a solution to this problem.  
Specifically, we propose a novel protocol for achieving Secure Multiparty Distributed Differentially Private (SM-DDP) computations on sensitive data. The protocol provides the guarantees of both SMC and DP. SMC is provided through Homomorphic Encryption (HE)~\cite{gentry2009fully} while DP is provided via Functional Mechanism (FM)~\cite{zhang2012functional}. 
An important characteristic of FM is that it injects noise into the feature matrices (\ie coefficients of objective function), which can be computed independently by each party in a multiparty computational environment. We explore this feature of FM and apply it to linear regression using our SM-DDP protocol, but it can be applied to the computation of statistical model function that allows independent calculation from the local statistics. We show that the accumulated noise in our protocol is still bounded and convergent by using the infinite divisibility property of Laplacian distribution~\cite{kotz2012laplace}. Finally, we evaluated SM-DDP protocol's computational efficacy on linear regression using two real-world datasets. 
We compare our results with the use of Centralized DP (CDP) in a multiparty setting as in Fig.~\ref{fig:com}. The intuition is that the distributed setting of DP (DDP), which is proposed in this paper, would cause a greater accuracy loss than the typical client-server setting of SMC systems. However, we show exactly same trade-off can be achieved using the SM-DDP protocol that is presented in Fig.~\ref{fig:framework}.
The extensive evaluation results indicate that the proposed SM-DDP protocol does not yield minimal computational overhead---less than a minute for 20 parties with 32 attributes and 10K samples. The individual parties obtain better accuracy than that would be obtained from a single party model. Finally, SM-DDP is scalable while providing security and privacy guarantees.

\noindent\textbf{Contributions:} In this paper, we summarize our contributions as follows: 
\begin{itemize}
\item We proposed a novel Secure Multiparty Distributed Differentially Private (SM-DDP) protocol to achieve secure and differentially private computations in distributed multiparty settings. This protocol can be applied to any statistical model function that allows the calculation of global model from the independent local statistics.
\item We implemented the SM-DDP protocol on linear models. We showed that SM-DDP allows parties \revision{to} compute regression model on pooled data while providing secure computation and differential privacy guarantees.


\item We showed that the accumulated noise in our protocol is bounded and convergent. This allows parties to build a model function, which offers the individual-level privacy against an untrusted data collector.  


\item We evaluated the performance of the proposed protocol using two different datasets. The results demonstrated that the parties compute the models in less than a minute while preserving the security guarantees of SMC and DP.

\end{itemize}

\noindent\textbf{Organization:} The rest of the paper is organized as follows: We present the related work in Section~\ref{sec:related}. In Section~\ref{sec:tech}, we give the technical preliminaries about SMC and DP methods that we utilized. Then, in Section~\ref{sec:linear}, background about regression analysis and specifically distributed calculation of linear regression is given. In Section~\ref{sec:framework}, a novel protocol for SM-DDP computation of a statistical model function $f$ and its application to linear regression is presented. Furthermore, we give the experimental results for the application of our protocol to linear regression in terms of accuracy, scalability, computational overhead, and security trade-offs in Section~\ref{sec:evaluation}. Finally, we discuss some of the related issues in Section~\ref{sec:discuss} and then we conclude the paper in Section~\ref{sec:Conclusion}. 
 
\section{Related Work} \label{sec:related}
There have been many works on the secure computation of linear regression over distributed databases~\cite{karr2005secure,du2004privacy,karr2009privacy,sanil2004privacy,hall2011secure}. In these, the threat model is considered as a third party that does not have access to data, but curious about it. However, one of the parties may want to release the model function after computing function securely, which still poses threats to the individuals~\cite{narayanan2008robust,ganta2008composition,el2013secure}. DP copes with this problem as it injects a certain amount of noise to the results of the queries to mask the individuals in the database. Indeed, there have been different works about the DP~\cite{dwork2006our,dwork2006calibrating,dwork2008differential,mcsherry2007mechanism} and particularly about differentially private linear regression~\cite{chaudhuri2011differentially,bassily2014private,duchi2013local,fredrikson2014privacy,jain2013differentially,zhang2012functional,smith2017interaction}. However, these works consider DP without SMC. Although they are useful, they only provide privacy guarantees that the output of queries does not carry information about the individuals. 

Approaches combining SMC and DP to provide both individual-level privacy and secure computation would be more secure. However, combining DP and SMC is not trivial; indeed, it is a rather challenging task since the application of centralized DP just after SMC in client-server settings would leak the model to \revision{an} untrusted data collector, which results in a privacy violation of individuals in the database. Applying distributed DP directly on the local data held by the parties is more secure, but \revision{if each user independently injects noise randomly,} it may lead to an excessive or \revision{uncontrollable} amount of accumulated noise at the data collector end. Recent works focused on combining SMC and DP~\cite{goryczka2013secure,clifton2013challenges,shi2011privacy}, but none of them focused on linear regression. \revision{As pointed in~\cite{zhang2012functional}, the main reason behind this is that the regression analysis involves an optimization problem, which makes it harder to control the required amount of noise, and if the data is also distributed among parties, that makes it much more difficult to control the privacy-accuracy trade-off introduced by DP.}\revision{In another relevant work}~\cite{pathak2010multiparty}, a combination of SMC and DP is proposed for aggregate classifiers. However, this approach injects the noise to the optimum model parameter. This resulted in excessive noise in the global model and significant loss in the accuracy. \revision{Particularly, the experimental evaluation shows that when the classifier is locally trained, the error rate obtained from locally trained classifiers is higher than the optimum error rates that could be obtained from a centralized approach.} However, in our work, we take a different approach from this work. We deploy FM~\cite{zhang2012functional}, which adds noise to local statistics, which provides the same model as the centralized approach. \revision{Lastly, even though a similar idea is proposed in~\cite{aono2015fast}, it is not analyzed in detail.}

\section{Linear Models} \label{sec:linear}
In this section, we start by introducing the linear models. We, then, show how to compute linear regression in a distributed fashion. 

\subsection{Background}
Assume a database $D$ consists of $n$ observations $\{x_i,y_i\}_{i=1}^n$, where $x_i$ is a vector of $d$ attributes (i.e., $x_i=(x_{i1},x_{i2},\dots,x_{id})$ and $y_i$ is a scalar response. The aim is to find a \emph{model function} $f: X \to Y$ that can predict $y_i \in Y$ as close as its actual value using the attributes $x_i \in X$. The type of the regression model is decided by the type of the model function. For instance, in linear regression, the model function is simply a straight line. Model function $f$ takes model coefficients $w=(w_1,w_2,\dots,w_d)$ and $x_i$ as inputs and outputs a prediction for the value of $y_i$. The deviations between predicted value and the actual response value are calculated through a \emph{loss function} $\ell: Y \times Y \to \mathbb{R}$. The global value of $w$ over the training data $D$ is calculated by the objective function. We denote the objective function by $\mathcal{L}$ and it is calculated as follows:
\begin{equation}
    \mathcal{L} (f,D)=\sum_{i=1}^n \ell(f(x_i,w),y_i).
\end{equation}

\subsection{Distributed Linear Regression}
Regression is a statistical approach that explores the relationships between a set of independent variables called \textit{attributes} and one dependent variable called \textit{response}. In regression, the relationship between the attributes and the response is modeled using a prediction function.

In linear regression, $L_2$-norm of the objective function (\ie $\ell(f(x_i,w),y_i)=(w \cdot x_i-y_i)^2 $) that is minimized in the matrix form as follows:
\begin{equation} \label{eqn:obj:linear}
w^*= \displaystyle\arg \min_{\substack{w}} \mathcal{L} (f,D)= \displaystyle\arg \min_{\substack{w}} \sum_{i=1}^m (w \cdot x_i-y_i)^2,
 \end{equation}
where $m$ is the number of tuples in the database. 
To calculate the regression in a distributed way, we represent the regression objective by minimizing with the \textit{Maximum likelihood Estimation} (MLE). MLE allows us to obtain the global solution of the Equation~\ref{eqn:obj:linear} as follows\footnote{A unique solution only exists if $ (\textbf{X}^\top \textbf{X})^{-1}$ is non-singular. In other cases, there are techniques for solving Equation~\ref{eqn:obj:linear}~\cite{mohriintroduction}; however, it is out of the scope of this paper.}:
\begin{equation}
w^*= (\textbf{X}^\top \textbf{X})^{-1}  \textbf{X}^\top \textbf{Y} .
\end{equation}

We characterize the model parameter $w$ of each party using three parameters:
\begin{center}
\begin{equation}
 \mathcal{P}_i= \textbf{X}_i^\top \textbf{X}_i, 
 \mathcal{V}_i= \textbf{X}_i^\top \textbf{Y}_i, 
 \mathcal{O}_i= \textbf{Y}_i^\top \textbf{Y}_i 
\end{equation}
\end{center}

Each party computes its \textit{local statistics} $<\mathcal{P}_i,\mathcal{V}_i,\mathcal{O}_i>$ and shares with other parties. Then, the global values of $\mathcal{P}$,$\mathcal{V}$ and $\mathcal{O}$ are  computed using the shared local statistics as follows:
$$\mathcal{P}= \textbf{X}^\top \textbf{X}=\begin{bmatrix}
           X^\top_{i_1}|...|X^\top_{i_n}
         \end{bmatrix} 
         \begin{bmatrix}
           X_{i_1} \\
           \vdots \\
           X_{i_n}
         \end{bmatrix} = \sum_{k=1}^n \textbf{X}_{i_k}^\top \textbf{X}_{i_k}= \sum_{k=1}^n \mathcal{P}_k$$ 
         
$$\mathcal{V}= \textbf{X}^\top \textbf{Y}=\begin{bmatrix}
           X^\top_{i_1}|...|X^\top_{i_n}
         \end{bmatrix} 
         \begin{bmatrix}
           Y_{i_1} \\
           \vdots \\
           Y_{i_n}
         \end{bmatrix} =  \sum_{k=1}^n \textbf{X}_{i_k}^\top \textbf{Y}_{i_k}= \sum_{k=1}^n \mathcal{V}_k$$ 

$$\mathcal{O}= \textbf{Y}^\top \textbf{Y}=\begin{bmatrix}
           Y^\top_{i_1}|...|Y^\top_{i_n}
         \end{bmatrix} 
         \begin{bmatrix}
           Y_{i_1} \\
           \vdots \\
           Y_{i_n}
         \end{bmatrix} = \sum_{k=1}^n \textbf{Y}_{i_k}^\top \textbf{Y}_{i_k}= \sum_{k=1}^n \mathcal{O}_k,$$
where $n$ is the number of parties in the collaboration. Using this, the global coefficients can be computed as follows:

\begin{equation}
w^*= (\textbf{X}^\top \textbf{X})^{-1}  \textbf{X}^\top \textbf{Y} = \mathcal{P}^{-1} \mathcal{V}.
\end{equation}

In order to calculate the error of the global function, we rewrite the objective function in Equation~\ref{eqn:obj:linear} in terms of the local statistics (\ie matrix form) as follows:

\begin{equation} 
\begin{split}
    \sum_{i=1}^m (w \cdot x_i-y_i)^2 &= (\textbf{X}w-\textbf{Y})^\top (\textbf{X}w-\textbf{Y})\\
    &=||(\textbf{X}w-\textbf{Y})||^2 \\
    &= w^\top \textbf{X}^\top \textbf{X} w- 2 w^\top \textbf{X}^\top \textbf{Y} + \textbf{Y}^\top \textbf{Y} \\
    &= w^\top \mathcal{P} w-2 w^\top \mathcal{V}+\mathcal{O},
    \end{split}
 \end{equation}
 where $||\cdot||$ denotes the Euclidean norm. We note that even though we do not need $\mathcal{O}$ to calculate the global coefficients, it is used for computing the error of the model.

\section{Technical Preliminaries} \label{sec:tech}
Preserving the privacy of the users and data is a long-studied problem in the area of cryptography~\cite{shi2011privacy,damgaard2012multiparty,dankar2015privacy,chaudhuri2011differentially,sanil2004privacy,dwork2006our}. As a result of these long-term studies, there are several theoretically well-studied tools that can be employed to protect the data and user privacy such as Secure Multiparty Computation (SMC)~\cite{damgaard2012multiparty} and Differential Privacy (DP)~\cite{dwork2008differential}. In this section, we introduce the essentials of the secure computation and differential privacy primitives to understand the implementation of SM-DDP algorithms. Particularly, we introduce Homomorphic Encryption (HE) to provide SMC and Functional Mechanism (FM) to provide DP guarantees.

\subsection{Secure Multiparty Computation}
SMC allows the computation of a function with multiple inputs from different users while keeping the users' inputs hidden from each other. For instance, each party $P_i$ in a $n$-party environment holds input $x_i$ learns nothing but the output $f(x_1,...,x_n)$ of a computation. In the literature, SMC schemes are mostly achieved via either the  Yao's garbled circuits~\cite{yao1982protocols} or Homomorphic Encryption (HE)~\cite{gentry2009fully}. In the following, we use HE to provide guarantees of secure computation. 

\vspace{3pt}

\noindent\textbf{Homomorphic Encryption (HE)-} HE provides an ability to evaluate the functions directly on the encrypted data while keeping the data confidential. The primary advantage of the HE is that it does not require any interaction between the parties other than the data exchange. That is, there is no additional communication complexity. However, it may introduce computational overhead on large plaintexts. Recent works improved its performance significantly by introducing new techniques like single instruction, multiple data (SIMD) operations~\cite{smart2014fully} or using different mathematical assumptions like learning with errors LWE~\cite{brakerski2012leveled,brakerski2014efficient} (see~\cite{2017arXiv170403578A} for a recent survey about HE).

\begin{figure}
\centering
\begin{tikzpicture}[>=stealth,thick, scale=0.4]
\node[] (A) {$m_1,m_2$};
\node[right=2cm of A] (B) {$f(m_1,m_2)$};
\node[above=2cm of A] (C) {$c_1,c_2$};
\node[right=2cm of C] (D) {$f(c_1,c_2)$};

\draw[shorten <=0cm,shorten >=1.5cm,-latex,line width=0.6mm] (C.0)--node[above left]{$Eval(...)$}(D.0);
\draw[shorten <=0cm,shorten >=0.5cm,-latex,line width=0.6mm] (A.100)--node[left]{$Enc_{pk}(...)$}(C.100);
\draw[shorten <=0cm,shorten >=0.5cm,-latex,line width=0.5mm] (D.-100)--node[right]{$Dec_{sk}(...)$}(B.-130);
\draw[shorten <=0cm,shorten >=1.8cm,-latex,line width=0.5mm] (A.0)--node[below left]{$f(...)$}(B.0);

\end{tikzpicture}
\caption{HE operations of encryption, evaluation, and decryption ($pk$ is the public key, $sk$ is the secret key, and $f$ is the function desired to be computed). \label{diagram}}
\end{figure}
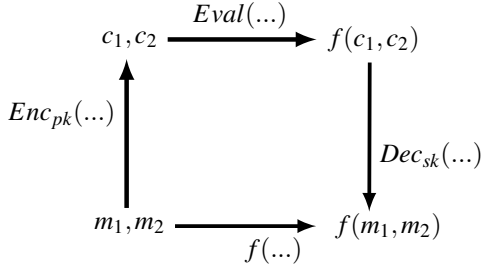

An HE scheme is primarily characterized by four operations: key generation ($KeyGen$), encryption ($Enc$), decryption ($Dec$), and evaluation ($Eval$). $KeyGen$ is the operation that is used to generate a secret and public key pair for the asymmetric version of HE or a single key for the symmetric version. $KeyGen$, $Enc$ and $Dec$ are similar to the ones used in conventional encryption schemes. However, $Eval$ is an HE-specific operation, which takes ciphertexts as input and outputs a ciphertext corresponding to a functioned plaintext. Fig.~\ref{diagram} illustrates a commutative diagram depicting the relationship among the four major operations. The simplified version of the diagram shows only one homomorphic encryption with two ciphertexts~\cite{gentry2014computing}.



\subsection{Differential Privacy (DP)} DP is a statistical disclosure control technique ensuring that the outputs of queries do not leak information about the individuals found in a dataset. It injects a certain amount of noise into the replies of the queries so that while it is not possible to infer an individual-level leak, the output of the query is still ``almost'' the same. In other words, query results of a data release algorithm for two closely similar data sets give the same answer. The formal definition of $\epsilon-$differential privacy is formulated as follows~\cite{dwork2014algorithmic}:

 \begin{definition}
A randomized algorithm $\mathcal{M}$ is $\epsilon$-differentially private if for all data sets $D$ and $D'$ differing on at most one element and all $S \subseteq Range(\mathcal{M})$,
\begin{equation}
Pr[\mathcal{M}(D) \in S] \leq \exp(\epsilon) \times Pr[\mathcal{M}(D') \in S],
\end{equation}
where $Range(\mathcal{M})$ shows all possible outputs of the function (query), $f$.
\end{definition}

The definition states that two adjacent sets $D$ and $D'$, which differs at most one element, act approximately the same against a query\footnote{The queries or functions correspond to the predictions in the statistical models.} defined by a given mechanism $M$. $\epsilon$ can be considered as the degree of the privacy guarantee and the amount of information which can be learned from a result of a single query is bounded by $\exp(\epsilon)$. Since $\epsilon$ is too small, its guarantee is preserved for consecutive queries. Differential privacy works on the release mechanism and does not modify data or the format of the data in any way. 

The parameter $\epsilon$, called \textit{privacy budget}, is the main parameter to tune the balance between privacy and accuracy. Decreasing $\epsilon$ increases the privacy guarantees while decreasing the accuracy. The common mechanism to control the amount of noise that needs to be added is \textit{Laplace Mechanism} (LM). In this \revision{case}, the noise is drawn from a Laplace Distribution. The probability density function of LM is as follows: 
\begin{equation} \label{eqn:laplace}
Lap(x|b)=\frac{1}{2b} exp \left(-\frac{|x|}{b}\right),
\end{equation}
for scale $b$ and center $0$. It is shown that LM preserves $\epsilon$-differential privacy~\cite{dwork2014algorithmic}.

\begin{definition}
Given any function $f: \mathbb{N}^{|\mathcal{X}|} \to \mathbb{R}^{k}$, the mechanism is a Laplace Mechanism $\mathcal{M}$ if: 
\begin{equation}
\mathcal{M}(x)=f(x)+ \eta,
\end{equation}
where $x \in \mathcal{X}$ and $\eta$ is a vector of independent and identically distributed random variables drawn from Lap($\Delta f / \epsilon$).
\end{definition}

In addition to the $\epsilon$, \textit{sensitivity} is another important parameter in DP to determine the optimum noise amount. It is defined as follows:
\begin{definition}
For a function $f:D \rightarrow R^k$, sensitivity of $f$ is
\begin{equation}
  \Delta f = \max_{D,D'} \parallel f(D)-f(D') \parallel
 \end{equation}
  for all $D, D'$ differing in at most one element.
\end{definition}
The sensitivity shows the maximum number of elements that can change in two different queries.

\vspace{3pt}

\noindent\textbf{Functional Mechanism (FM)-} FM is an algorithm that is used to provide differential privacy guarantees for a set of linear models~\cite{zhang2012functional}. It is an extension of the Laplace Mechanism. The goal of the algorithm is injecting the noise to the polynomial coefficients of a model's objective function. This is accomplished with the mechanism of \textit{objective perturbation}~\cite{chaudhuri2011differentially}. The optimization of the noisy objective function gives new model parameters that ensure the $\epsilon$-privacy of each element in a database. Algorithm~\ref{alg:func_mech}~\cite{zhang2012functional} presents the functional mechanism.

\begin{algorithm}
\caption{\cite{zhang2012functional} Functional Mechanism ($D$,  $\mathcal{L}$, $\epsilon$)}\label{alg:func_mech}
\begin{algorithmic}[1]
\Require Let $\mathcal{L} (f,D)=\displaystyle \sum_{j=1}^J  \sum_{\phi \in \Phi_j} \sum_{i=1}^n \lambda_{\phi_i} \phi (w)$
 \State Set $\Delta=2 \displaystyle \max_{\substack{w}} \sum_{i=1}^n ||\lambda_{\phi_i}||_1 $ 
\For{each $j \in \{0,...,J\}$}
    \For{each ${\phi \in \Phi_j}$}
        \State $\lambda_{\phi}=\sum_{i=1}^n \lambda_{\phi_i}+Lap(\frac{\Delta}{\epsilon})$ \Comment{$noise~inject$}
    \EndFor
\EndFor
\State Compute new $w^*= \displaystyle\arg \min_{\substack{w}} \mathcal{L} (f,D)$ \Comment{$optimize$} \\
\Return $w^*$
\end{algorithmic}
\end{algorithm}

\begin{figure*}
        \centering
          \includegraphics[width=.75\textwidth,trim= 4cm 1.5cm 8cm 2cm]{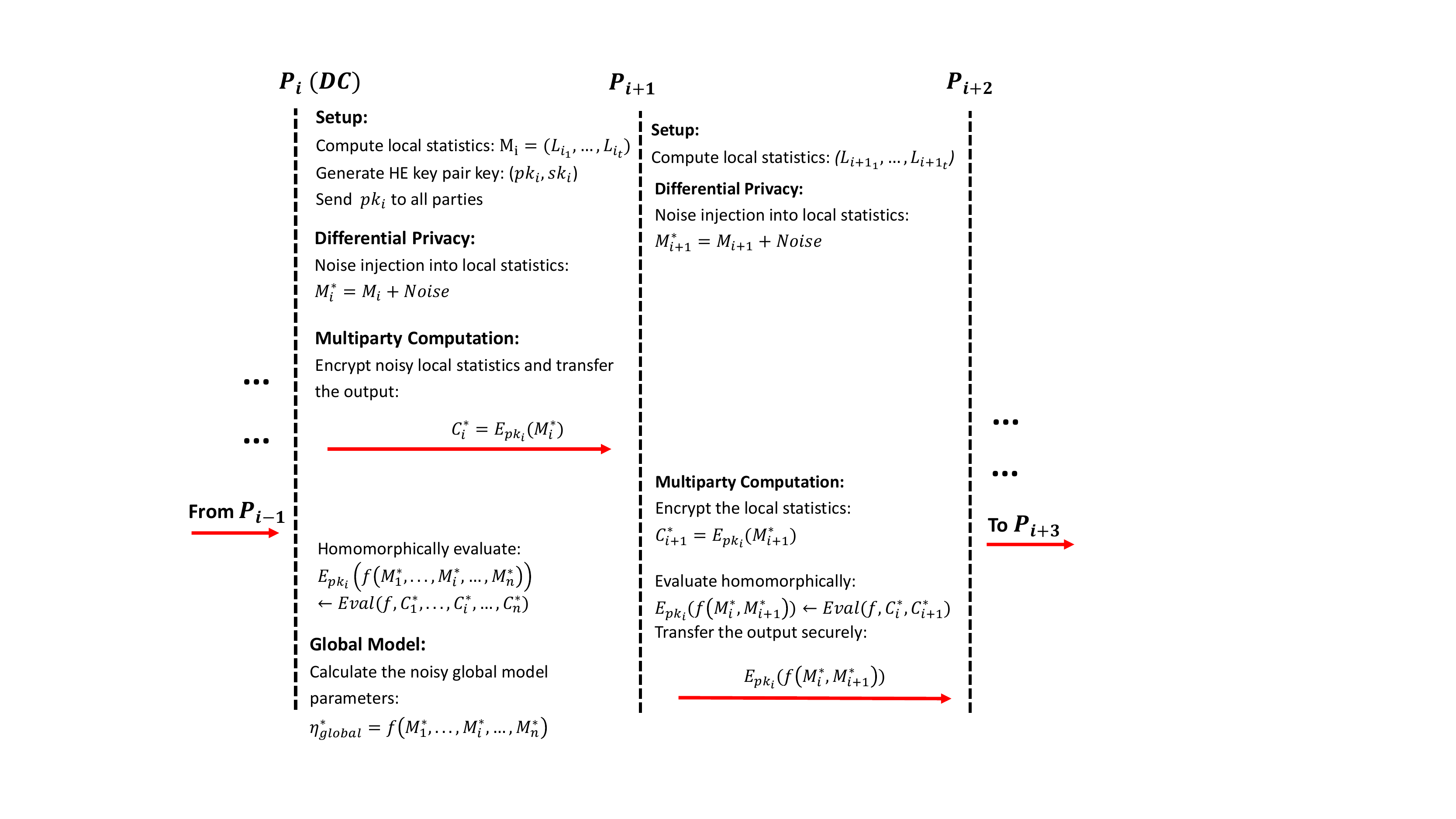}
          \caption{Secure Multiparty Distributed Differentially Private (SM-DDP) protocol for the computation of a linear model coefficients. The parties create a ring topology and the Data Collector (DC) initiates the protocol. The protocol can be applied to any statistical model function that allows independent calculation of local statistics.    \vspace{-10pt}
          \label{fig:framework}}
\end{figure*}

As illustrated in Algorithm~\ref{alg:func_mech}, FM takes a dataset $D$,  the polynomial representation of the objective function $L$, and the privacy budget $\epsilon$ as inputs and it returns the differentially private model coefficients $w^*$. It firstly injects noise drawn from a Laplacian distribution ($Lap(\frac{\Delta}{\epsilon})$) into all the coefficients $\lambda_{\phi_i}$ of the polynomial representation of the objective function and then the optimization is performed using noisy coefficients. It is shown that it satisfies $\epsilon$-differential privacy~\cite{zhang2012functional} \ie the predictions using $w^*$ does not leak any information about an individual in the database data. For example, if we have a quadratic objective function in the matrix form of $w^\top \mathcal{P} w+w^\top \mathcal{V}+\mathcal{O}$, where $\mathcal{P}$, $\mathcal{V}$, and $\mathcal{O}$ are the coefficients of the polynomial representation of the objective function. FM firstly injects noise into the coefficients, which results in $w^\top \mathcal{P}^* w+w^\top \mathcal{V}^*+\mathcal{O}^*$. Then, the optimization problem (i.e., $w^*= \displaystyle\arg \min_{\substack{w}} \mathcal{L} (f,D)$) is solved using $\mathcal{P}^*$, $\mathcal{V}^*$, and $\mathcal{O}^*$.

\section{Secure and Differentially-private Distributed Computations}\label{sec:framework}

\begin{algorithm*}
\caption{Computation of Linear Regression using SM-DDP protocol
}\label{distributed}
\begin{algorithmic}[1]
\Require Each party holds a database in the format of $D_i=(x_i,y_i)_{i=1}^n$ i.e., horizontally partitioned 
\Statex \hspace{0.4cm} The global privacy budget $\epsilon$.
\Ensure The differentially private global regression model of $D=\cup_{i=1}^n D_i$ 

\Algphase{Setup: Runs at the party $P_i$ (DC)}
\State $({pk}_i, {sk}_i) \leftarrow KeyGen()$ \Comment{generate the key pair of HE}
\State $\eta_{max}$, $\eta_{min} \leftarrow ComputeMinMax(D)$ \Comment{calculate the global max and min of each attribute via~\cite{ccetin2015depth}}
\State $\Delta \gets 2 (d+1)^2$ \Comment{calculate the global sensitivity, $d$ is the number of attributes}

\Algphase{Secure Regression Protocol: each party $P_j$ runs locally}
\hspace{0.4cm} \Require Received aggregate statistics for all previous parties as:
\hspace{0.4cm} \Statex $\xi$: ${E_{{pk}_i}(\sum_{k=1}^{j-1}\mathcal{P}_k^*})$ 
\hspace{0.4cm} \Statex $\kappa $: ${E_{{pk}_i}(\sum_{k=1}^{j-1}\mathcal{V}_k^*})$ 
\hspace{0.4cm} \Statex $\delta $: ${E_{{pk}_i}(\sum_{k=1}^{j-1}\mathcal{O}_k^*})$ 
\vspace{0.3cm}
    \State  $D_j^{norm}  \gets (D_j- \eta_{min})/(\eta_{max}-\eta_{min})$ \Comment{perform min-max normalization}

   \State $\mathcal{P}_j \gets \textbf{X}_j^\top \textbf{X}_j $, $\mathcal{V}_j \gets \textbf{X}_j^\top \textbf{Y}_j $, and $\mathcal{O}_j \gets \textbf{Y}_j^\top \textbf{Y}_j $ \Comment{compute local statistics}
   \State  $\epsilon_i \gets \alpha \epsilon$ \Comment{compute its share from the global privacy budget }
    \State $({\mathcal{P}_j^*},{\mathcal{V}_j^*},{\mathcal{O}_j^*}) \gets FM . NoiseInject({\mathcal{P}_j},{\mathcal{V}_j},{\mathcal{O}_j})$ \Comment{apply FM noise injection}
    \State $C_{j}^*=\big({E_{{pk}_i}(\mathcal{P}_i^*}),E_{{pk}_i}({\mathcal{V}_i^*}),E_{{pk}_i}({\mathcal{O}_i^*})\big)$ \Comment{perform encryption}
    \Statex \Comment{add its own encrypted local statistics to the received aggregate statistics}
     \State ${E_{{pk}_i}(\sum_{k=1}^{j}\mathcal{P}_k^*}) \gets {E_{{pk}_i}(\mathcal{P}_j^*})+ \xi $ 
     \State ${E_{{pk}_i}(\sum_{k=1}^{j}\mathcal{V}_k^*}) \gets {E_{{pk}_i}(\mathcal{V}_j^*})+ \kappa $ 
     \State ${E_{{pk}_i}(\sum_{k=1}^{j}\mathcal{O}_k^*}) \gets {E_{{pk}_i}(\mathcal{O}_j^*})+ \delta $ 
    \State \textbf{Send(~}${E_{{pk}_i}(\sum_{k=1}^{j}\mathcal{P}_k^*}),~ {E_{{pk}_i}(\sum_{k=1}^{j}\mathcal{V}_k^*}),~ {E_{{pk}_i}(\sum_{k=1}^{j}\mathcal{O}_k^*}) $\textbf{~)} to $P_{j+1}$ \Comment{send updated aggregate statistics to the next party.}
    \vspace{0.1cm}
    
\Algphase{Reconstruction: runs at the party $P_i$ (DC)}
\hspace{0.4cm} \Require Received aggregate statistics for all parties as:
\hspace{0.4cm} \Statex $\xi $: ${E_{{pk}_i}(\sum_{k=1}^{n}\mathcal{P}_k^*})$ 
\hspace{0.4cm} \Statex $\kappa $: ${E_{{pk}_i}(\sum_{k=1}^{n}\mathcal{V}_k^*})$ 
\hspace{0.4cm} \Statex $\delta $: ${E_{{pk}_i}(\sum_{k=1}^{n}\mathcal{O}_k^*})$ 
\vspace{0.3cm}

\State  $\mathcal{P}^* \gets D_{{sk}_i}\Big({\xi}\Big)$ \Comment{acquire the cleartext}
\State  $\mathcal{V}^* \gets D_{{sk}_i}\Big({\kappa}\Big)$ \Comment{acquire the cleartext}
\State  $\mathcal{O}^* \gets D_{{sk}_i}\Big({\delta}\Big)$ \Comment{acquire the cleartext}
\State  $({\mathcal{P}^*},{\mathcal{V}^*},{\mathcal{O}^*}) \gets FM . Optimize({\mathcal{P}^*},{\mathcal{V}^*},{\mathcal{O}^*})$\Comment{apply optimization}

\State  $w^* \gets {\mathcal{P}^*}^{-1} {\mathcal{V}^*}$ (i.e., $ w^*= \displaystyle\arg \min_{\substack{w}} w^\top \mathcal{P}^* w+w^\top \mathcal{V}^*+\mathcal{O}^*$)  \Comment{compute the global parameters}
\State $ Err \gets {w^*}^\top {\mathcal{P}^*} {w^*}+ {w^*}^\top \mathcal{V}^*+\mathcal{O}^*$ 
\State \textbf{Publish(~}$w^*,~Err$\textbf{~)} to all parties.
\Statex \Comment{Use of Model:}
\State $f(x_i,w^*) \gets \sum_{i=1}^n x_i {w^*}_i$ for an input $x_i \in \textbf{X}_i$ \Comment{computes the normalized predictions}
\State  $y_{pred} \gets f(x_i,w^*)(\eta_{max}-\eta_{min})+\eta_{min}$\Comment{perform de-normalization to get actual values}
\end{algorithmic}
\end{algorithm*}

In this section, we propose a novel protocol for secure multiparty distributed and differentially private (SM-DDP) computations through the use of homomorphic encryption (HM) and functional mechanism (FM). We evaluate its application to linear regression and discuss its extension to the logistic regression that can be used in supervised classification.

Consider $n$ parties $P_1,\dots,P_n$, where each has private horizontally distributed database $D_1,\dots,D_n$. Each database consists of a certain number of tuples in the format of $t_i=(x_i,y_i)$. The parties would like to jointly build a linear model of the pooled database $f(D)$, where $D=\cup_{i=1}^{n} D_{i}$ so that the security guarantees of both SMC and DP are preserved. Before running the protocol, each party in the collaboration agrees on the function to be computed and compute a collection of local statistics $M_i=(L_{i_1},\dots,L_{i_t})$.  We assume the linear model can be computed using the local statistics generated by each party independently i.e., $\eta_{global}=f(M_1,\dots,M_i,\dots,M_n)$. We define the guarantees and goals of our protocol as follows:
\begin{itemize}
    \item \textit{Individual privacy:} No information leaks about the individuals in the private databases held by the parties, \ie tuples $t_i$ is not leaked.
    \item \textit{Data privacy:} Information about the statistics of the data does not leak in the databases held by the parties, \ie the statistics about the data $M_i$ is not leaked. 
    \item \textit{Correctness:} The parties receive the correct output of the model.
\end{itemize}
We note that using SMC only would violate the individual privacy while using DP only violates the data privacy. In our combined protocol, we achieve individual privacy through FM and data privacy through HE and since all operations in the protocol are deterministic, the correctness is satisfied by design. We note that we assume there is a secure channel between parties to exchange messages.

Fig.~\ref{fig:framework} illustrates our protocol to be able to perform SM-DDP computations. It is initiated by one of the parties called \textit{data collector} (DC). In the setup phase, DC generates a key pair $({pk}_i,{sk}_i)$ and computes its own local statistics $M_i$ independent from other parties. Then, in the next phase, DC applies DP by injecting (adding) noise drawn from a random distribution that satisfies $\epsilon$-differential privacy into its local statistics. The encryption of the noisy local statistics is transmitted to the next party $P_{i+1}$. The next party $P_{i+1}$ also computes its local statistics and injects noise into them. The result is encrypted with $pk_i$ and the function is evaluated homomorphically with the inputs of parties $P_i$ and $P_{i+1}$. The protocol is continuous in the same way, where parties are located in a ring topology. At the final step, the securely evaluated function result is used by the party $P_i$ which decrypts it with $sk_i$. In the end, $P_i$ reveals the differentially private global model.

\vspace{-.3cm}
\subsection{Case Study: Linear Regression} \label{subsec:case}
In this subsection, we show how to compute linear regression using our protocol proposed in Fig.~\ref{fig:framework}. Particularly, we use functional mechanism shown in Algorithm~\ref{alg:func_mech} by splitting it into two parts: $NoiseInject()$ and $Optimize()$. In $NoiseInject()$, the noise drawn from Laplacian distribution (Equation~\ref{eqn:laplace}) is injected into each coefficient of the polynomial representation of the objective function. Then, in $Optimize()$, the optimization problem of the objective function is solved by applying regularization and spectral trimming introduced in~\cite{zhang2012functional} in order to avoid unbounded noisy objective function. Moreover, in FM, it is assumed that $\sqrt{\sum_{i=1}^{d}x_{id}^2}  \leq 1$. Therefore, a secure maximum computation is performed to calculate $\eta_{min}$ and $\eta_{max}$ in setup phase of Algorithm~\ref{distributed}, where $\eta_{min}$ (resp. $\eta_{max}$) is vector consists of global minimum (resp. maximum) of each attribute. Before applying FM, each party normalizes its database using the global maximum and minimum values. This guarantees that the local sensitivity of the parties is always same as the global sensitivity as we focus on the horizontally distributed data.

Algorithm~\ref{distributed} illustrates the computation of linear regression algorithm using the protocol presented in Fig.~\ref{fig:framework}. In linear regression,
the global model is calculated by simply aggregating locally calculated noisy statistics. While aggregating the local statistics, the noise of each party is aggregated as well. Therefore, it is necessary to make sure the final model will not violate $\epsilon$-differential privacy nor cause an unbounded noise. Particularly, the noise is injected to each coefficient as follows:
\begin{equation}
 {\mathcal{P}_i}^*=\mathcal{P}_i+Lap\Big(\frac{\Delta}{\epsilon_i}\Big). 
 \end{equation}
Then, when DC computes the global model, the local statistics are summed up as follows:
\begin{equation}
 \mathcal{P}^*=\sum_{i=1}^n {\mathcal{P}_i}^*= \sum_{i=1}^n \bigg(\mathcal{P}_i+Lap\Big(\frac{\Delta}{\epsilon_i}\Big)\bigg)= \mathcal{P}+ \sum_{i=1}^n Lap\Big(\frac{\Delta}{\epsilon_i}\Big).
 \end{equation}
Moreover, $\mathcal{V}^*$ and $\mathcal{O}^*$ can be computed similarly. In all $\mathcal{P}^*$, $\mathcal{V}^*$, and $\mathcal{O}^*$, the noise term is $\sum_{i=1}^n Lap\big(\frac{\Delta}{\epsilon_i}\big)$. In order to make sure that the accumulated noise is also Laplacian distribution, we use the following theorem.


\begin{theorem}
Let $Y$, $Y_1$, $Y_2$... be non-degenerate and symmetric i.i.d. random variables with variance $\sigma^2>0$, and let $\nu_p$ be a geometric random variable with mean $1/p$, independent of the $Y_i$'s. Then, the following statements are equivalent (Proof is given in~\cite{kotz2012laplace}): \\
(i) Y is stable with respect to geometric summation, i.e., there exist constants $a_p>0$ and $b_p \in \mathbb{R} $, such that 
\begin{equation} \label{eqn:lap_sum}
  a_p \sum_{i=1}^{\nu_p} (Y_i+b_p)=Y \quad \forall p \in(0,1)
 \end{equation}
 (ii) Y possesses the Laplace distribution with mean zero and variance $\nu_2$. Moreover, the constants $a_p$ and $b_p$ must be of the form: $a_p=\sqrt{p}$, $b_p=0$
\end{theorem}

From the theorem above, a Laplace distribution can be calculated by summing up several Laplace distributions in a certain form. In other words, the sequence of partial sums, $a_p \sum_{i=1}^{\nu_p} (Y_i+b_p)$ converges to a Laplace distribution under beta-distributed $a_p$. 
We addressed requirements of the theorem in
Algorithm~\ref{distributed} by multiplying the noise distribution of local parties with a number drawn from the geometric distribution i.e., $a_p \sum_{i=1}^n Lap\big(\frac{\Delta}{\epsilon_i}\big)$, where $a_p$ is a geometric random variable.


\section{Performance Evaluation} \label{sec:evaluation}
In this section, we give the experimental results for the application of our SM-DDP protocol to linear regression. Table~\ref{table:notations} presents the notations used throughout the experiments. We first demonstrate how we set the parameters that are introduced in the distributed setting. Particularly, the success probability of the geometric random variable $p$ in Equation~\ref{eqn:lap_sum} and $\alpha$ introduced in Algorithm~\ref{distributed} is investigated. After experimentally tuning these two parameters, we test the final protocol with a different dataset without random sampling directly as it is collected. During evaluation, we focus on the following questions: (\rmnum{1}) Can we obtain a differentially private global linear regression model from differentially private local statistics? (\rmnum{2}) Does our approach support up to 100 parties? (\rmnum{3}) How long does it take to complete the protocol? (\rmnum{4}) Does it guarantee the security and privacy of both data and individuals? We analyzed and discussed each of these questions in Sections~\ref{subsec:accuracy}-\ref{subsec:security}.

\vspace{3pt}

\noindent\textbf{Dataset-} We used two real-world datasets to evaluate the algorithms of our protocol. Both datasets include highly sensitive data. The first dataset is \textit{Integrated Public Use Microdata Series} (IPUMS)~\cite{ipums}. It contains 370K decennial census records of people living in the US with 14 attributes, 7 of which are demographic information and the rest are working hours per week, the number of years residing in the current location, the number of children, the number of automobiles, and the annual income. The attributes are used to predict the \textit{annual income} of a person. The second dataset is the warfarin dataset collected by the International Warfarin Pharmacogenetics Consortium (IWPC)~\cite{international2009estimation}. The dataset contains clinical and genetic data of patients to predict the stable therapeutic dose of warfarin. Clinical data includes demographics, background, and phenotypic attributes. Genetic data includes genotype variants of CYP2C9 (*1, *2 and *3) and VKORC1 (one of seven single nucleotide polymorphisms in linkage disequilibrium). 21 sites in 9 countries and four continents contributed to the dataset. We used a subset of this dataset wherein patient samples include no missing attributes. Overall, we used 1400 complete patient samples from seven medical institutions. We used IPUMS dataset to experimentally set the parameters of our protocol and we tested the final protocol with the IWPC dataset, where each party corresponds to a medical institution in the dataset.

\vspace{3pt}

\noindent\textbf{Evaluation Metrics-} We applied stratified cross validation to split the dataset into training and test sets. To evaluate the model's prediction accuracy, we used \textit{Mean Squared Error} (MSE) as it is a commonly used metric for linear regression analysis. It is calculated as $\frac{1}{n} \sum_{i=1}^n (\hat{y_i}-y_i)$, which gives the average of the squared errors between actual ($y_i$) and predicted ($\hat{y_i}$) values in $n$ data samples. The lower values of MSE shows better predictions. Finally, it is worth mentioning that all the experiments show 100 independent runs and their average is reported in this work.

\begin{table}
\centering
\caption{Abbreviations and notations used in experiments}
\label{table:notations}
\begin{tabular}{|p{.7cm}|p{3.3cm}|l|} 
\hline 
Notation     & Description                                                 & Range                                \\ \hline \hline 
DDP          & Distributed Differential Privacy                            & -                                    \\ \hline
NoDP         & No Differential Privacy                                     & -                                    \\ \hline 
CDP          & Centralized Differential Privacy                            & -                                    \\ \hline 
$\epsilon$   & global privacy  budget                                      & \{0.1,0.2,0.4,0.8,1.6,3.2,6.4,12.8\} \\ \hline 
$\epsilon_i$ & local privacy budget                                        & $\epsilon_i=\alpha \epsilon$         \\ \hline 
$\alpha$     & local privacy ratio \ie $\alpha = \epsilon_i / \epsilon$    & \{1,10,100\}                         \\ \hline 
p            & success probability of  the geometric random variable, $a_p$ & \{0.1,0.5,0.9\}                      \\ \hline 
n            & number of parties                                           & {[}1,100{]}                          \\ \hline 
L            & number of levels in HElib                                    & \{4,6\}                              \\ \hline 
$nslots$     & number of slots in HElib                                    & calculated by HElib                  \\ \hline 
s            & minimum of $nslots$                                         & \{$8^2,16^2,24^2,32^2,40^2$\}      \\ \hline  
\end{tabular}
\end{table}

\noindent\textbf{Experimental Setup-} To evaluate the computational overhead, we used open-source HE library (HElib)~\cite{helib}, which implements BGV homomorphic cryptosystem~\cite{brakerski2012leveled} and we ran experiments on 16-core Intel Xeon CPU at 1.90 GHz running Linux Server. In BGV, a prior level $L$ should be set before initiating the computation. In addition to the level $L$, HElib also has a parameter $nslots$ which defines a number of slots for the utilization of SIMD techniques~\cite{smart2010fully,smart2014fully}. HElib allows encrypting multiple messages at one time through its SIMD features by packing the messages into the independent slots of an array. We note that the parameter $L$ affects not only the number of allowed homomorphic operation but also all the other timings and the key size. Therefore, the parameter $L$ should be optimized so that the minimum $L$ is set without failure of the decryption. To do so, we first calculated the table of a number of homomorphic operations for each level $L$ and we used the minimum level for each number of the party.

Furthermore, in our experiments, the data encrypted is the local statistics \ie not the raw data. The size of the local statistics is considered the same for all the parties. The homomorphic operation computed for linear regression is the element-wise matrix addition. To take advantage of HElib library SIMD features, we converted matrices into arrays and the parameter of minimum number for $nslots$ was set to the length of the array for each statistics. This prevents data loss during the conversion. We did not utilize any multi-threading technique during our experiments to see the lower bound of the performance of our protocol. Thus, our results are lower bound and can be improved with the use of any multi-threading technique.
\vspace{5pt}
\subsection{Accuracy Analysis} \label{subsec:accuracy}

\begin{figure}
        \centering
          \includegraphics[width=.4\textwidth,trim=1cm 0cm 17.5cm 0cm]{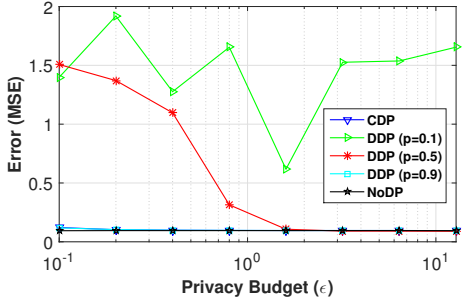}
          \caption{Tuning $p$ \label{fig:dp_1}. Variation of error is tested for several values of $p$. As a result, $p=0.1$ is not stable or convergent; $p=0.5$ is convergent, but error is much higher than CDP for especially small $\epsilon$ values. Hence, we chose $p=0.9$ as the best case.}
\end{figure}

\begin{figure}
        \centering
          \includegraphics[width=.4\textwidth,trim=1cm 0cm 17.5cm 0cm]{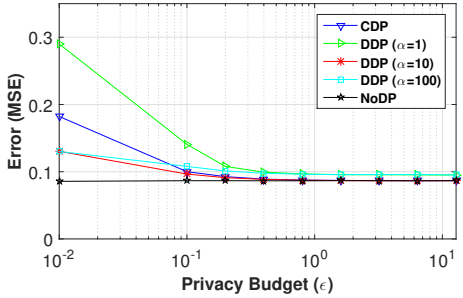}
          \caption{Tuning $\epsilon_i$. Variation of error is tested for several values of local privacy budget $\epsilon_i$ for $\alpha=\epsilon_i / \epsilon$. For $\alpha=1$, error is too high for small $\epsilon$ values. For $\alpha=10$, error is lower than CDP and and it converging to the value as NoDP. For $\alpha=1$, error is low, but it converges to a value higher than NoDP. Hence, we chose $\alpha=10$ as the best case.\label{fig:dp_2}}
          \vspace{-7.5pt}
\end{figure}

\begin{figure}
        \centering
          \includegraphics[width=.4\textwidth,trim=1cm 0cm 17.5cm 0cm]{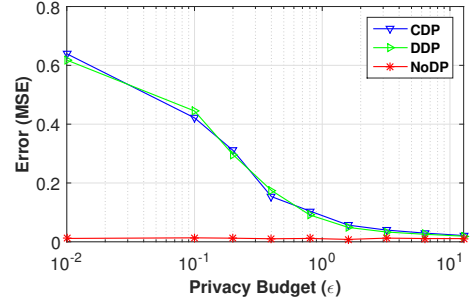}
          \caption{A real test: Warfarin dataset with 7 parties with $\epsilon_i=n\epsilon$ and $p=0.9$. Exactly the same trade-off as the centralized differential privacy is obtained. \label{fig:warfarin}}
           \vspace{-7.5pt}
\end{figure}

\begin{figure}
        \centering
          \includegraphics[width=.4\textwidth,trim=1cm 0cm 17.5cm 0cm]{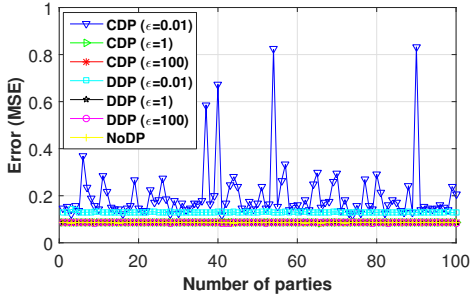}
          \caption{Impact of number of parties in the collaboration for $\epsilon_i=n\epsilon$ and $p=0.9$. \label{fig:dp_3}}
           \vspace{-10pt}
\end{figure}

We evaluate the accuracy-privacy trade-off of distributed evaluation of differential privacy on linear regression. 
Specifically, we compare our results with the centralized approach. In \textit{Centralized Differential Privacy} (CDP), the accuracy of the regression depends only on the global privacy budget $\epsilon$. However, in \textit{Distributed Differential Privacy} (DDP), each party has its own local privacy budget $\epsilon_i$ and DDP is applied independently by each party. We note that this is a particular property of FM. In FM, data is first normalized and the optimum noise amount is only determined by the number of the attributes which is same for all parties. Therefore, the size and the range of the local statistics are same for all the parties; it does not depend on the number of tuples in the local database. Since all parties are identical, we choose the same local privacy budget $\epsilon_i$ for all the parties. Finally, in our fist three experiments (Fig.~\ref{fig:dp_1},~\ref{fig:dp_2}, and~\ref{fig:dp_3}), we used IPUMS dataset and split it into parties using random sampling methods. In the last experiment, we used IWPC dataset for accuracy evaluation. We split the dataset based on the given medical institutions (See Fig.~\ref{fig:warfarin})

The first set of experiments was conducted to analyze the optimum value of $p$, which is a parameter of geometric random variable $a_p$ given in Equation~\ref{eqn:lap_sum}. In theory, $a_p$ is required to obtain a Laplace distribution in the global model, thereby it is required to be able to satisfy $\epsilon$-differential private model. To present the impact of the parameter $p$ on the accumulated global noise, we kept the party number constant for several values of $p$ and various $\epsilon$ values ($\epsilon_i=\epsilon$). To do so, each party multiplies the noise drawn from Laplace distribution with a random variable $a_p$, which is a geometric random variable with success probability $p$. We compared the error rates of CDP, DDP, and NoDP algorithms in terms of MSE.

Fig.~\ref{fig:dp_1} illustrates the error and privacy budget trade-off for various values of $p$. We varied $p$ from $\{0.1,0.5,0.9\}$. We found that DDP with $p=0.1$ does not converge to a value while increasing the value of $\epsilon$. However, $p=0.5$ and $p=0.9$ converges to the same value as NoDP as it is desired and when $p$ is $0.9$, it gives similar results to CDP. In the sequel, we tuned $p=0.9$ and used it in our experiments.

In the second set of experiments, we were interested in finding the optimal local privacy budget $\epsilon_i$ for a predetermined global privacy budget. In other words, we assume all parties agree on a global privacy budget according to the sensitivity of the dataset, which was indeed calculated by the number of attributes. We denote the ratio of local privacy budget to the global privacy budget as $\alpha$, \ie $\alpha=\epsilon_i / \epsilon$. We first tried the value of $\alpha$ less than $1$, the result of DDP was much worse than CDP. This is because smaller $\epsilon_i$ means more noise injected locally by each party than the centralized approach. This noise decreases the accuracy significantly. Therefore, we changed $\alpha$ from $\{1,10,100\}$ and compared the results with CDP and NoDP mechanisms. The results are presented in Fig.~\ref{fig:dp_2}. We found that if $\alpha$ is the number of parties, which is $10$ in this experiment, the plot gets closer to CDP and the error is converging to NoDP, which is the desired case. Therefore, in the rest of experiments, we set $\alpha=n$, where $n$ is the number of parties.

\begin{figure*}
        \centering
          \includegraphics[width=.95\textwidth,trim=0cm 1cm 1cm 0cm]{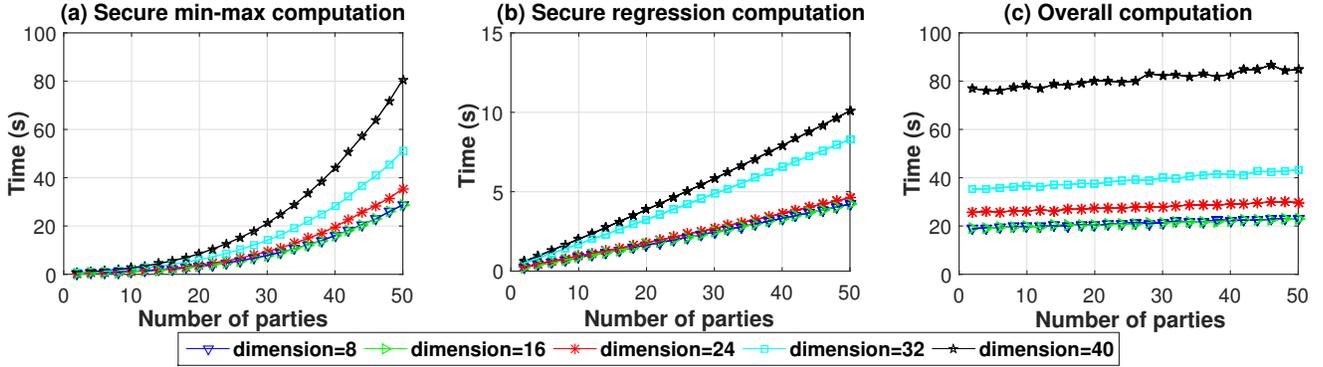}
          \vspace{10pt}
          \caption{Performance evaluation of SM-DDP computations of linear regression algorithm.\label{fig:performance}}
          \vspace{-5pt}
\end{figure*}

So far, we tuned the parameters of our approach experimentally. Now, in our last experiment, we evaluated the efficiency of our protocol using the dataset (IWPC dataset) collected from multi sources. We applied DP locally on each party's dataset and calculated the global model and error. Our goal was to see the feasibility of our approach in a real case and test the feasibility of our approach. 

In this experiment, we set $\epsilon_i=n\epsilon$, $p=0.9$ as we found in earlier experiments. We compared the performance of CDP, DDP, and NoDP algorithms. Fig.~\ref{fig:warfarin} shows MSE rates for varying $\epsilon$. We found that the same trade-off with CDP can be achieved by applying DP while training the classifiers locally. We note the DDP is also converging to the error of NoDP when $\epsilon$ approaches infinity as desired. 

\subsection{Scalability Analysis}

In this set of experiments, we evaluated the scalability of our proposed protocol. We set $\epsilon_i= n \epsilon$, where $n$ is the number of parties; as we found $\alpha=n$ is optimum and for a different number of parties, we split the dataset into the number of parties ($n$) by using random sub-sampling. Then, each party applies DP locally, but we note that the pooled dataset is still the same.

Laplace distribution is infinitely divisible~\cite{kotz2012laplace}. Therefore, the accumulated error of global model should not be affected by the number of parties. We ran the analysis for some users ranging from 1 to 100 and present the results in Fig.~\ref{fig:dp_3}. The results demonstrated an interesting point, which is when $\epsilon=0.01$, even though CDP is not stable, DDP is. On the other hand, when $\epsilon$ is 1 or 100, the error rate stays the same even for 100 parties. This means our protocol is scalable even for 100 parties.

\subsection{Computational Overhead Analysis}
In this subsection, we evaluate the computational overhead of linear regression presented in Algorithm~\ref{distributed}. We found that DP algorithms do not introduce computational overhead. Therefore, we only evaluate the computational overhead of our SMC algorithm, which consists of three main parts: Key generation of HE, min-max, and regression computation.

Fig.~\ref{fig:performance} shows the computation time for different dimension sizes. Fig.~\ref{fig:performance}a presents the time for secure computation of finding global min-max of each attribute. It increases quadratically with the number of parties. However, this algorithm runs at the setup phase, so it is performed before initiating the computations. There are two interesting results worth to note. First, the time of secure regression computation increases linearly as a number of parties in the collaboration increases, but with a different slope for dimension, which is illustrated in Fig.~\ref{fig:performance}. The reason for the linear increase is that the number of encryptions and homomorphic evaluations are directly scaled by the number of parties in the group.
Second, similar results hold for the overall computation time (see Fig.~\ref{fig:performance}c), but as a minor change since the key generation time shifts the lines in the y-axis and also increases the scale. However, similar to the secure min-max computation, the execution of the key generation algorithm does not require all parties in the group to be online since it occurs in the setup phase. On the other hand, we also note that size of the local database of each party does not have an impact on the total computational time since parties only share the local statistics, which is dependent on the attribute size, instead of the raw data. As can be seen in Fig.~\ref{fig:performance}c, the overall computation of the protocol including both offline and online phases for 20 parties with 32 attributes and 10K samples is less than a minute. Hence, our SM-DDP protocol yields minimal computational overhead.
\vspace{-5pt}
\subsection{Security and Privacy Analysis} \label{subsec:security}
In this section, we discuss the security and privacy guarantees of SM-DDP protocol given in Fig.~\ref{fig:framework}. As all the communication among the parties is encrypted, the security of the algorithm is simply reduced to the security of underlying HE scheme. A leak can occur \revision{only} if DC is corrupted since the data is encrypted using the public key generated by DC. However, \revision{even in this case, DC will only obtain the noisy local statistics, not the raw data, and at the end of the protocol,} DC has only control over the aggregated data while reconstructing the global model and it can not know which party contributed to the result. While the protocol is running, the view of all the other parties consists of homomorphically encrypted data. Therefore, if the given homomorphic encryption scheme is semantically secure, the parties can not distinguish the corresponding plaintexts. So, the computation is private even in the presence of an honest, but curious adversary model presented in~\cite{goldreich2009foundations}. Therefore, data privacy is preserved. 

On the other hand, we both showed theoretically (Section~\ref{subsec:case}) and experimentally (Fig.~\ref{fig:warfarin}), a differentially private global model can be obtained through the locally applied DP. Therefore, it is not possible that an untrusted data collector can infer information about the individuals.
Furthermore, the collaboration comes with a price as the local parties used $\epsilon_i$ instead of $\epsilon$. Therefore, the local privacy guarantee is decreased by $\alpha$ (\ie $\epsilon_i$ is increased by $\alpha$), even though the global model's guarantee is still the same, meaning that data privacy against an untrusted DC is still preserved and the local privacy guarantee is important only if the underlying SMC is bypassed. Finally, since we set $\alpha$ as the number of parties in the collaboration, each party should take this into consideration while deciding on the global privacy budget.  


\section{Discussion} \label{sec:discuss}
\vspace{-5pt}
The preceding analysis showed how to achieve secure multiparty computation and differential privacy in distributed settings focusing on linear regression on horizontally distributed data. That is, parties do not see each others' inputs and further can not infer individuals' data from the final constructed model. A limitation of our algorithm is that we assume parties do not collaborate to learn a target party's input. \revision{However, if the party that generates the key pair conspires with the parties that are neighbors of a target in the ring topology, the noisy local statistics ($\xi $, $\kappa $, $\delta $) of the victim can be extracted.} More generally, this is known as \textit{active corruption}, where the data collector is an active attacker and has control over the other corrupted parties. \revision{Our protocol in Fig.~\ref{fig:framework} achieves only a collusion threshold of 1, but the distributed DP algorithm that we present here can easily be adapted to work with recent solutions in SMC such as~\cite{damgaard2012multiparty}, which is secure in the presence of an active adversary corrupting up to $n-1$ of the $n$ parties. To extend our work with these more secure SMC schemes, it suffices to use the noisy output of the functional mechanism instead of using the local statistics directly as input to the underlying SMC algorithm.}

In our evaluation, we used HElib, an implementation of the fully homomorphic operation, to compute generic results. It supports both addition and multiplication; however, \revision{while computing the linear regression coefficients, we only used the addition operation.} The performance of secure computation can be improved by using other libraries such as Paillier cryptosystem~\cite{paillier1999public}, which is \revision{only} additively homomorphic cryptosystem.

Finally, our algorithms can be easily extended to other algorithms such as logistic regression in a supervised classification setting. In logistic regression, each party independently computes a score vector $u_i$ and information matrix $\mathcal{I}_i$. Instead of injecting noise to the local statistics as in linear regression, noise can be injected into $u_i$ and $\mathcal{I}_i$ vectors. However, the optimization of objective function differs in logistic regression as it requires several iterations. Fortunately, there exist some techniques that let implementing the iterations for computing the secure multi-site logistic regression~\cite{el2013secure}. Combining this secure multi-site logistic regression algorithm with FM would solve this issue. We defer the detailed application of this method to future work.

\section{Conclusion}\label{sec:Conclusion}
\vspace{5pt}
In this work, we have proposed a novel Secure Multiparty Distributed  Differentially Private (SM-DDP) protocol to achieve private computations in a multiparty environment as an application in linear regression. Using homomorphic encryption and functional mechanism, we first presented a  protocol to provide the guarantees of secure multiparty computation and differential privacy. Then, we built the algorithms that would allow distributed parties to compute a global model while preserving the privacy of their data and individuals found in the dataset. Any statistical model function that can be independently calculated by sharing the local statistics of the parties can be computed through this protocol. Finally, we evaluated the performance of the proposed protocol on two datasets, namely, warfarin dose and budget predictions. Our findings show that a party can achieve individual-level privacy via our proposed protocol for distributed differential privacy, which is independently applied by each party in a distributed fashion. Moreover, the experiment results demonstrated that the proposed SM-DDP protocol is both feasible and scalable that is its computational overhead is minimal and overall computation time is sub-linear with the number of parties. Indeed, SM-DDP protocol provides security and privacy guarantees while being feasible and scalable.
Our future work will extend the algorithms outside the linear models and investigate the accuracy and performance trade-offs of other algorithms. \revision{We are also planning to compare the performance of Laplacian mechanism used in FM with other DP mechanisms such as Exponential Mechanism~\cite{mcsherry2007mechanism} and  Sample-and-aggregate~\cite{nissim2007smooth}.}
\section*{Acknowledgment}
\revision{This work was partly supported by US NSF-CAREER-CNS-1453647 and Army Research Laboratory under Cooperative Agreement Number W911NF-13-2-0045 (ARL Cyber Security CRA).  
%
The views and conclusions contained in this document are those of the authors and should not be interpreted as representing the official policies, either expressed or implied, of the Army Research Laboratory or the U.S. Government. The U.S. Government is authorized to reproduce and distribute reprints for Government purposes notwithstanding any copyright notation here on.}

\newpage
\bibliographystyle{IEEEtran}
\bibliography{references}

\end{document}